# Performance Comparison Analysis of ArangoDB, MySQL, and Neo4j: An Experimental Study of Querying Connected Data

Johan Sandell
Stockholm University
josa5094@student.su.se
johan.sandell8@gmail.com

Einar Asplund
Stockholm University
eias6938@student.su.se
einar.asplund@gmail.com

Workneh Yilma Ayele
Stockholm University
workneh@dsv.su.se

Martin Duneld
Stockholm University
xmartin@dsv.su.se

**Abstract**

*Choosing and developing performant database solutions helps organizations optimize their operational practices and decision-making. Since graph data is becoming more common, it is crucial to develop and use them in big data with complex relationships with high and consistent performance. However, legacy database technologies such as MySQL are tailored to store relational databases and need to perform more complex queries to retrieve graph data. Previous research has dealt with performance aspects such as CPU and memory usage. In contrast, energy usage and temperature of the servers are lacking. Thus, this paper evaluates and compares state-of-the-art graphs and relational databases from the performance aspects to allow a more informed selection of technologies. Graph-based big data applications benefit from informed selection database technologies for data retrieval and analytics problems. The results show that Neo4j performs faster in querying connected data than MySQL and ArangoDB, and energy, CPU, and memory usage performances are reported in this paper.*

## 1. Introduction

These days, it has become more pervasive to deal with data-rich relationships, such as social media apps (Fernandes & Bernardino, 2018, p. 1; Robinson et al., 2015, pp. 11-24). Since graph data is becoming more common, it follows that choosing and developing performant database solutions that can handle big data, allowing for big data analytics and machine learning with complex relationships efficiently, is essential (Robinson et al., 2015, pp. 11-24; SAS White Paper, 2012, p 1, Tsirogiannis et al., 2010, p. 231; Xu et al., 2012, p. 1954). Furthermore, graph databases are needed by organizations of diverse domains, such as companies that put effort into developing large systems, for instance, healthcare, online business solutions, financials, and communications, as their query response time is in milliseconds (Kaliyar, 2015, p. 787).

The governance and management of databases can significantly impact organizations and societies since inefficiencies can lead to higher energy consumption (Tsirogiannis et al., 2010, p. 231; Xu et al., 2012, p. 1954). In recent years, companies have provided services that cannot be achieved efficiently using relational databases (Jouili & Vansteenberghe, 2013; Phaneendra & Reddy, 2013, pp. 3386-3391). An alternative is to use a graph database (Fernandes & Bernardino, 2018, p. 1).

It is essential to develop and use database infrastructure solutions that can handle big data with complex relationships to understand better and provide services with high and consistent performance (Robinson et al., 2015, pp. 11-24; SAS White Paper, 2012, p 1, Tsirogiannis et al., 2010, p. 231; Xu et al., 2012, p. 1954). Compared to relational databases, it is difficult for developers to identify the particular type of use cases for which each graph database is most suitable (Pokorný, 2015, pp. 68-69). The performance can also vary across different graph database technologies depending on the size of the graph and how well-optimized a given tool is for a particular task (Pokorný, 2015, pp. 68-69).

Previous research overlooks some of the performance aspects between graph databases and relational databases, such as CPU usage, memory usage, power usage, and temperature of the servers, as presented in Chapter 2. Another research gap is discussed by Tsirogiannis et al. (2010, p. 231) and Xu et al. (2012, p. 1954), which measured the power usage between databases. However, their studies were conducted a decade ago and may be outdated. Different graph database technologies continuously update and release their computational and data storage engines. Neo4j has released a new major version yearly and more minor improvements approximately twelve times a year (Neo4j Inc., 2023).

Similarly, ArangoDB usually releases one to two new major versions yearly (ArangoDB Inc., 2023).

MySQL usually releases less frequently, with the last major release in 2016 and more minor updates at least two to three times yearly (Oracle, 2023a; Oracle, 2023b). This means that older comparisons are at risk of needing to be updated. Thus, the problem that has given rise to this work is the need for more knowledge on how state-of-the-art graphs and relational databases perform in different aspects to allow developers to make more informed choices for their applications.

This study aims to answer the following research question: *How do state-of-the-art-graph database and relational database technologies compare from a performance perspective regarding time taken to query, CPU usage, memory usage, power usage, and server temperature?* We used an experimental study to compare the querying performance of connected data running on ArangoDB, MySQL, and Neo4j.

This paper is organized into six chapters. The chapters are the Background and related research, Experimental setup, Results, and Discussions, and we conclude the final chapter, Conclusions, and future directions.

## 2. Background and related research

The main topic of this study is to evaluate how different state-of-the-art database technologies compare with each other when querying connected data. Additionally, this study compared how graph database management systems differ from relational database technologies from a performance perspective regarding the time taken to query, CPU usage, memory usage, power usage, and server temperature. The central topics discussed in the background are relational databases, NoSQL databases, and benchmarks.

### 2.1. Databases

In modern days, it has become more common to deal with data-rich relationships, such as social media apps (Ayele et al., 2017, p. 1; Fernandes & Bernardino, 2018, p. 1; Robinson et al., 2015, pp. 11-24; SAS White Paper, 2012, p 1). This data type with complex relationships can be called a graph or a graph structure (Das et al., 2020, p. 1; Kaliyar, 2015, p. 788; Pokorný, 2015). Das et al. (2020, p. 1) describe a graph as a structure that consists of nodes, referred to as points, vertices, or entities, where edges, referred to as relations, connect the nodes, and each connection between the nodes have properties, referred to as attributes.

### 2.2. Relational databases

Relational databases are not primarily designed to deal with highly connected or graph data, such as data from social media platforms. Instead, relational databases are suited to support codifying tabula structures and paper forms (Robinson et al., 2015, pp. 11-24). Over the years, the relationships between data evolved, and developers had to work around the issue of relationships since the data had become more connected (Phaneendra & Reddy, 2013, pp. 3386-3391; Robinson et al., 2015, pp. 11-24). What made the relational databases struggle were the ad hoc modeling and data relationships that mimicked the real world (Robinson et al., 2015, pp. 11-24). This issue worsened over time since the outlier data multiplied (Fernandes & Bernardino, 2018, p. 1; Mpinda et al., 2015, pp. 87-89; Phaneendra & Reddy, 2013, pp. 3386-3391). Another problem with relational databases is that the relationships have become complex joins between tables (Robinson et al., 2015, pp. 11-24). Thus, much complexity is added, and queries have become computationally expensive (Robinson et al., 2015).

### 2.3. NoSQL databases

NoSQL database technologies were created to deal with increasing NoSQL data by storing the data as either a key-value pair, column-oriented database, or document database. Similarly, the issue of storing, retrieving, and analyzing increasingly growing connected NoSQL data is solved with the emergence of graph database technologies with increased performance (Robinson et al., 2015, pp. 11-24). Graph database technologies are built based on graph structures having nodes and edges represented in the data structure (Pokorný, 2015, p. 1); thus, the data is not stored tabularly (Kaliyar, 2015, p. 789). Das et al. (2015, pp. 2–3) and Pokorný (2016, p. 1) identified a key difference between graph and relational databases. Graph databases like Neo4j, an open-source implemented in Java, do not use fixed schemas (Das et al., 2015, p. 4).

Graph databases are needed by organizations of diverse domains, such as companies that put effort into developing large systems, as their query response time is in milliseconds (Kaliyar, 2015, p. 787). Also, there are many application areas where graph technologies are in demand. For example, data management of IoT systems as they have several connected sensors (Ueta et al., 2016, p. 299). Graphs can represent connected real-world entities, such as IoT sensors and social media users. Graphs can also represent semantic information, such as knowledge graphs using RDF data. Knowledge graphs have garnered significant attention from both industries and institutions in scenarios that require exploiting diverse, dynamic, large-scale data collections (Hogan et al., 2021). Pokorný (2016, s. 1).

## 2.4. Benchmarks

Benchmarking is used to evaluate existing software and hardware by measuring aspects of a system. For evaluating the performance of database technologies or any other software performance, benchmarks evaluate hardware performance while manipulating databases stored in computing machines (Alyas et al., 2023; Batra & Tyagi, 2012; Jouili & Vansteenberghe, 2013; Mpinda et al., 2015). Free benchmarking software, such as Novabench (Novabench Inc., 2023) or Speccy (Piriform Software Ltd., 2023), exists for this purpose. There also exist benchmarks tailored to measure the performance of databases, such as benchAnt (BenchAnt, 2023). On the other hand, some authors used their custom-built benchmarks, for example, Jouili & Vansteenberghe (2013).

## 2.5. Previous and related research

This section presents a list of previous and related research on the comparison of different database technologies:

- ➢ Tsirogiannis et al. (2010, p. 231) studied how database systems affect and fluctuate servers' energy efficiency when running queries. It was found that the power consumption between operators could vary as much as 60% (Tsirogiannis et al., 2010).
- ➢ Batra and Tyagi (2012) compared Neo4j with the relational database MySQL by comparing the response time from database queries. Results showed that Neo4j outperformed MySQL (Batra & Tyagi, 2012).
- ➢ Xu et al. (2012, p. 1954) developed a tool for evaluating the power consumption of databases and optimizing queries. Results implied that conventional query optimizers often ignored energy-efficient candidate queries (Xu et al., 2012, p. 1954).
- ➢ Using benchmarking, Cheng et al. (2019) conducted an experiment on the performance of graph and relational database technologies. Their results indicate relational database technologies have more advantages in query processing time and resource consumption than graph technologies.

The most recent work on the comparison of database technologies was done by Alyas et al. (2023). Alyas et al. (2023) compared the performance of MySQL and graph databases in terms of memory usage and execution time running on a premises cloud server, and they confirmed that graph databases are more suitable for working with corresponding data. In summary, most previous research works were done at least ten years ago and do not measure performance such as energy consumption, temperature fluctuations, memory, and CPU scores.

## 2.6. Motivation and Summary

We reviewed literature about relevant state-of-the-art previous research, and this study builds upon chosen previous works related to the research question. Previous studies conducted similar research by benchmarking different databases. The results from the previous research papers have brought up new interesting questions to answer, which this study builds upon. For example, there needs to be more in the previously listed research to measure some aspects of hardware performance, such as CPU, memory, and power usage between different databases. As graph data becomes more common, there is a need to evaluate and compare how performant different databases are at handling data with complex relationships (Tsirogiannis et al., 2010, p. 231; Xu et al., 2012, p. 1954). Such evaluations and comparisons can be conducted using a benchmark (Oxford Learner's Dictionary, 2023).

## 3. Experimental setup

In this paper, we used a quantitative experimental method to measure the performance of the databases. The performance data was collected using the scientific method described by Johannesson and Perjons (2014, p. 84) through an experiment where some workloads (queries) were tested against a benchmark tool for the respective database, and the results were observed. Performance data for a baseline was defined as the machine's performance when no query was running. Each measurement was done according to the following procedure. First, the baseline data was collected. Directly after, the benchmark was running in combination with the query on the chosen database technology.

### 3.1. Computational tool and software used

We used a computer with the specifications illustrated in Table 1. We also used Python for statistical analysis of the results using JuPyter,

Numpy, Pandas, Matplotlib, SkLearn, Scipy, and Scikt_posthocs.

| Machine Specifications | |
|---|---|
| Computer model | MSI GF63 Thin 11UC |
| Operating system | Windows 11 home 64-bit |
| CPU | Intel Core i5 (11th gen) @ 2.70GHz, Tiger Lake 10nm Technology. Number of cores: 6 Number of threads: 12 |
| GPU | NVIDIA GeForce RTX 3050 Laptop GPU |
| Memory | RAM: 32 GB DDR4 |
| | SSD: WDC PC SN540 SDDPNPF-512G-1032, Western Digital. Interface/ connector: SATA 1.5 Gb/s with 22-pin SATA connector. Capacity: 476 GB |

**Table 1. Machine specification that will be used.**

### 3.2. Chosen dataset

We used graph data from an open dataset (Open Graph Benchmark, 2023), "ogbl_biokg", published for research as the data source to populate the databases, as similarly used by Bellini & Nesi (2018). The chosen dataset contains medical graph data, and Figure 1 illustrates the relationships between diseases, drugs, side effects, functions, and proteins. We chose "ogbl_biokg" because it contains multiple types of nodes (entities), which enabled us to create a more realistic scenario of having many tables in a relational database and allowing for measuring performance.

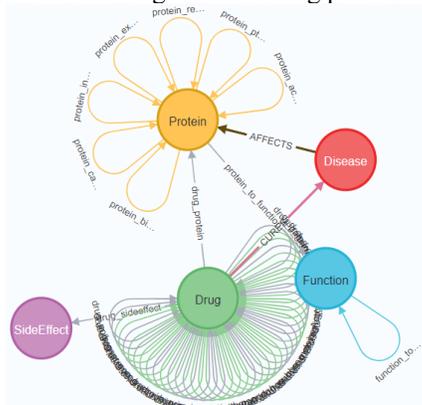

**Figure 1. Schema of the dataset in Neo4j.**

### 3.3. Formulating query

Since ArangoDB, MySQL, and Neo4j use different languages for querying, the same query could not be used across all database technologies. Instead, the query had to be written three times in AQL for ArangoDB, SQL for MySQL, and Cypher Query Language for Neo4j. Appendix A contains all formulated queries.

### 3.4. Chosen benchmark

There are several different benchmarking software, for example, Novabench (2023), BenchAnt (2023), and Speccy (Piriform Software Ltd., 2023). Previous research also shows that authors developed benchmark software to evaluate performances, such as Jouli & Vansteenberghe (2013). In this research, we used Novabench (Novabench, 2023) to measure performance measures, such as CPU usage, memory usage, power usage, and temperature, and the corresponding variables used are CPU Score, Memory Score, Avg. Low Load Temp, Avg. High Load Temp, and Avg. Energy Usage respectively. The time taken to execute queries (Time to Query) was also calculated by respective database technology. Novabench is commonly used to test computer performances (Sabolski et al., 2014, p. 59; Novabench, 2023).

### 3.6. Data analysis method

This study analyzed data using an ANOVA test, which, according to Ostertagova & Ostertag (2013) and Hae-Young (2013), is a statistical procedure concerned with comparing means of several samples. ANOVA is an analysis method characterized as a t-test with extensions to compare more than two independent groups of observations (Denscombe, 2014, pp. 260-262; Ostertagova & Ostertag, 2013). When performing an ANOVA procedure, it is required that the observations are independent of one another. The observations in each group should come from a normal distribution, and the population variances in each group are the same (Ostertagova & Ostertag, 2013). The following hypothesis was formulated: H0 - There is no significant difference in performance between the database technologies. H1 - There is a significant difference in performance between the database technologies.

The assumption of normality was also tested by checking the asymmetry and tail-heaviness, also known as skewness and kurtosis, which is a good indicator of normal data according to George and Mallery (2010), Hair et al. (2010) and Bryne (2010). The assumption of homogenous variances was tested with Bartlett's test because it generally produces good results (National Institute of Standards and Technology, 2012). However, Bartlett's test is sensitive to deviations from the normal distribution. In cases where the collected performance data diverges from the normal distributions, Levene's test is more appropriate and was used instead (National Institute of Standards and Technology, 2012).

Since ANOVA is sensitive to normally distributed data and homogeneity of variances, variations of ANOVA which are less sensitive to such requirements were used to improve validity by providing greater statistical power in case some of the assumptions do not hold (Borg & Westerlund, 2012; Delacre et al., 2019). For example, Welch's ANOVA gives better results when the variances are not homogenous, and the Kruskal-Wallis H-test gives better results when data is not normally distributed (Borg & Westerlund, 2012; Delacre et al., 2019). Further, post hoc tests were conducted to determine which database technologies differ (Borg & Westerlund, 2012). Scheffe's test was used where possible since it is the most flexible and conservative test (Ostertagova & Ostertag, 2013). However, according to Borg and Westerlund (2012), Dunn's test is better suited to non-parametric tests such as Kruskal Wallis H-test. Hence, it will be used in these cases instead.

## 4. Results

The preprocessing of collected data, descriptive statistics of performance measurements, test assumptions of normality and homogeneity of variances, ANOVA results, and post-hoc tests are presented in this chapter.

### 4.1. Preprocessing of the data and sample size

A sample size of 30 is desired when performing the ANOVA test (Ross & William, 2017) for reasonable statistical power of 80% (Ellis, 2010) as a rule of thumb (VanVoorhis & Morgan, 2007).

A few outliers were observed in the collected data by visualizing them using boxplots; see Appendix B - outliers. Removing these outlier data points is not always a feasible strategy. Besides, it is still being determined whether these outliers are incorrectly measured. Borg & Westerlund (2012, pp. 474-476) claim that ignoring outliers on the assumption that they are wrong might lead to studying an arranged reality. On the other hand, according to Borg & Westerlund (2012), ANOVA is also sensitive to outliers. Thus, it was decided to conduct all statistical tests with outliers present and verify results by conducting all tests twice with outliers replaced with central values.

### 4.2. Descriptive statistics of performance measurements

Table 2 illustrates descriptive statistics of performance measurements: standard deviation and mean values of Time to Query, CPU score, Memory score, Average low load temperature, and Average energy usage.

| Aspect | Statistics | Time to query [second] | CPU score [Score] | Memory score [Score] | Avg. low load temp [Celsius] | Avg. high load temp [Celsius] | Energy usage [Watt] |
|---|---|---|---|---|---|---|---|
| ArangoDB | Mean | 0.6037 | -26.9333 | 0.1000 | 3.3000 | 2.2333 | 3.3648 |
|  | SD | 0.1810 | 23.6783 | 1.0939 | 2.6412 | 8.2574 | 2.8661 |
| MySQL | Mean | 0.2803 | -31.8333 | -2.3333 | 9.6333 | 3.8000 | 4.0021 |
|  | SD | 0.0325 | 16.9869 | 0.9942 | 2.9882 | 6.0423 | 3.6155 |
| Neo4j | Mean | 0.0203 | -36.5333 | -0.0333 | 4.4000 | 3.9333 | 2.4822 |
|  | SD | 0.0066 | 22.7213 | 0.7649 | 4.2149 | 10.1606 | 4.2163 |

**Table 2. Summary of descriptive statistics.**

### 4.3. Test assumption of normality

ANOVA is sensitive to the data following a normal distribution (Borg & Westerlund, 2012). In cases where the data does not follow a normal distribution, Borg & Westerlund (2012) recommend using a variation of ANOVA, such as the Kruskal-Wallis H-test instead. We calculated the skewness and kurtosis values to confirm the normality assumption for the performance data collected. There needs to be more consensus on skewness and kurtosis cutoff rate. However, George and Mallery (2010), Hair et al. (2010), and Bryne (2010) argue that data is considered to have a normal distribution if skewness is between -2 to +2 and kurtosis is between -7 to +7. Thus, we considered the intervals suggested for the kurtosis and skewness by the authors mentioned above. Also, all aspects except for Avg. Low Load Temp can be considered to follow a normal distribution see Appendix C - symmetry, tail-heaviness, and normality assumption.

### 4.4. Test assumption of homogeneity of variances

ANOVA is also sensitive to the data having homogenous variances, which needs to be tested (Borg & Westerlund, 2012). Not all collected performance data for each aspect follows a normal distribution. Bartlett's test was used when data followed a normal distribution because of greater statistical power (National Institute of Standards and Technology, 2012). However, Bartlett's test is sensitive to deviations from the normal distribution. Levene's test is more suitable in such situations. Therefore, Levene's test was used when data did not follow a normal distribution (National Institute of Standards and Technology, 2012). It can be concluded from the test of the assumption of homogeneity of variances that the assumption of equal variances holds for all aspects except Time to Query.

### 4.5. ANOVA results

According to Borg & Westerlund (2012), it is not advisable to conduct ANOVA when the required assumptions of normality and homogeneity of variances do not hold, as the resulting P-value might be invalid. When the assumption of normality does not hold, alternative variations of ANOVA should be used, such as the Kruskal-Wallis H-test with some loss of statistical power. We used Welch's ANOVA as the assumption of homogeneity of variances did not hold, but the assumption of normality held following the recommendation by Delacre et al. (2019). As can be seen from Table 3, the aspects of Time to query and Memory score showed statistically significant results. Additionally, the aspect of low load temperature was on the verge of being significant, as it had a P-value less than 0.05 but test-statistics less than the critical value. No other aspects showed significant results.

| Aspect | df-total | df-between | df-within | P-value | Statistic value | Critical value |
|---|---|---|---|---|---|---|
| Time to query | (2.0, 40.2210) | 2.0 | 40.2219 | $1.8268 \cdot 10^{-35}$ | 1053.2939 * | 3.23 |
| CPU score | (2.0, 87.0) | 2.0 | 87.0 | 0.2481 | 1.4164 * | 3.10 |
| Memory score | (2.0, 87.0) | 2.0 | 87.0 | $3.0077 \cdot 10^{-17}$ | 60.8037 * | 3.10 |
| Avg. low load temp | 29 | N/A | N/A | $1.5416 \cdot 10^{-9}$ | 40.5808 ** | 42.557 |
| Avg. high load temp | (2.0, 87.0) | 2.0 | 87.0 | 0.6803 | 0.3868 * | 3.10 |
| Energy usage | (2.0, 87.0) | 2.0 | 87.0 | 0.2666 | 1.3420 * | 3.10 |

\* F-statistic and ** Chi-square
**Table 3. Results of ANOVA**

### 4.6. Post-hoc tests

We carried out Dunn's test to determine which technologies differed in Time to query, and Table 4 illustrated the resulting P-value. The obtained P-values, which are less than 0.05, indicate that all the database technologies differed significantly. As can be seen from Figure 2, the order from fastest to slowest is Neo4j, MySQL, and ArangoDB.

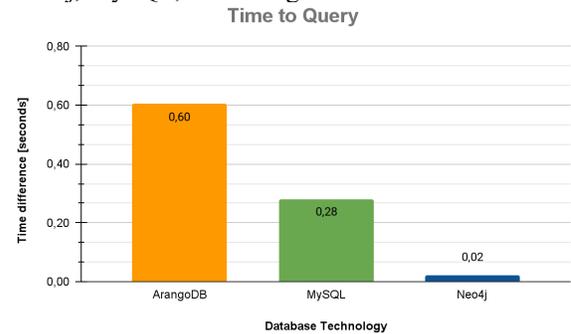

**Figure 2. Average execution time difference of the query for the database technologies.**

|  | ArangoDB | Neo4j | MySQL |
|---|---|---|---|
| ArangoDB | 1.0000e+00 | 4.8219e-17 | 0.000226 |
| Neo4j | 4.8219e-17 | 1.0000e+00 | 0.000003 |
| MySQL | 2.2558e-04 | 2.7217e-06 | 1.0000e+00 |

**Table 4. P-values from Scheffe's test for time to query, with chosen P=0.05.**

We carried out Scheffe's test to determine which database technologies differed in Memory score, and Table 5 presents the resulting P-values of the test. The P-values of MySQL compared with Neo4j and MySQL compared to ArangoDB is less than 0.05, which indicates significant difference amongst each other. On the other hand, the P-value of Neo4j compared to ArangoDB is greater than 0,05, which means there is no significant difference between the two. Figure 3 illustrates the average memory score difference for each database technology, clearly supporting that ArangoDB and Neo4j performed way better than MySQL.

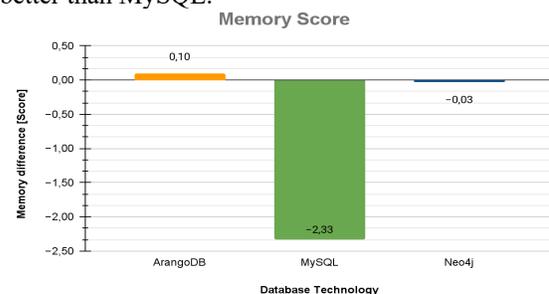

**Figure 3. Average memory score difference of the query for the database technologies.**

|  | ArangoDB | Neo4j | MySQL |
|---|---|---|---|
| ArangoDB | 1.0000e+00 | 8.6574e-01 | 8.5730e-15 |
| Neo4j | 8.6574e-01 | 1.0000e+00 | 1.0499e-13 |
| MySQL | 8.5730e-15 | 1.0499e-13 | 1.0000e+00 |

**Table 5. P-values from Scheffe's test for memory score with chosen P=0.05.**

## 5. Discussions

Graph or connected data has become pervasive with the growth of digital technologies and social media platforms (Ayele et al., 2017, p. 1; Fernandes & Bernardino, 2018, p. 1; Robinson et al., 2015, pp. 11-24; SAS White Paper, 2012, p 1). There are several graph databases and technologies, such as Neo4j. On the other hand, database technologies, such as MySQL, are designed to deal with SQL-based tabular data (Robinson et al., 2015, pp. 11-24), while the developers of ArangoDB designed it to support both NoSQL-based and SQL-based databases such as graph and document data (ArangoDB Inc., 2023). This paper evaluated the performance of SQL-based and NoSQL-based database technologies, Neo4j, ArangoDB, and MySQL, on connected data.

The query performance of Neo4j was significantly faster than the two other database technologies. Also, MySQL performed faster than ArangoDB. The result aligns with previous research, except for the performance difference between MySQL and ArangoDB. Previous findings confirm that graph database technologies, such as Neo4j, are faster than relational technologies regarding query execution time in general (Alyas et al., 2023; Batra & Tyagi, 2012; Li & Manoharan, 2013, pp. 15-19). On the other hand, ArangoDB performed well during the early testing and exploration phase when the query was less complex, despite the claim by ArangoDB Inc (2023) that it is optimized for connected data. The result might be due to several other reasons. For example, McColl et al. (2014, pp. 11-18) concluded that some graph databases might suffer from the issue of scaling. Thus, ArangoDB may have scaling issues, which must be well-established using empirical research.

Regarding Memory and CPU usage, Cheng et al. (2019) showed that relational database technologies perform better than graph databases. The memory performance difference could be because relational databases use on-disk rather than on-memory database processing models. On the contrary, the results suggest that graph databases are less memory-intensive than relational databases. The comparison of the database technologies on CPU usage showed no significant difference. Similarly, Cheng et al. (2019) also showed that relational databases performed inconsistently and inefficiently when the data increased in volume and complexity. The measurements of Memory Score in this study are less consistent for MySQL than for ArangoDB and Neo4j, which supports this.

We did not find a significant difference in the average energy usage compared to previous research, as discussed by Tsirogiannis et al. (2010, p. 23) and Xu et al. (2012, p. 1954). We can assume that database technologies have become more aware of the importance of energy optimization, or further experiments to substantiate this claim have to be made. Nevertheless, although it is statistically insignificant, ArangoDB and Neo4j had a lower mean energy usage value than MySQL. It still cannot be excluded that the included outliers have impacted the statistical tests and thus also affected the validity of the results by inflating or deflating mean values. Even if we calculated changes in computational performance, the reason for outliers could be unpredictable changes in background processes.

To discern whether the identified outliers impacted the results, see Appendix B - outliers, ANOVA was conducted twice on imputed data. The outliers were replaced with the means and the median of each database technology performance aspect. In both cases, a difference was observed in the average load temp, which followed a normal distribution and gave a statistically significant result in ANOVA, showing that Neo4j and ArangoDB differed from MySQL. No other changes in the result could be observed in the other performance measurement aspects, strengthening the validity and credibility of these results.

Since database technologies use different query languages, it becomes difficult to create identical queries. In particular, AQL was challenging due to the complex syntax and a lack of experience in AQL. The results' validity, generalizability, credibility, and extensibility could be affected negatively since the query may be partially the same as in other languages. Finally, the database technologies compared are representatives of existing database types. For example, Neo4j and ArangoDB are NoSQL database technologies where ArangoDB supports both SQL and NoSQL types while Neo4j supports connected data particularly. On the other hand, MySQL is a representative of SQL-based database technologies.

## 6. Conclusions and future directions

We aimed to evaluate the performances of state-of-the-art database technologies. The performance evaluation included not only the speed of the execution of a complex query but also CPU Usage, Memory Usage, energy usage, and temperature of connected information. The results from the ANOVA and post-hoc tests show significant differences between the database technologies, which answered our research question by rejecting the null hypothesis, H0, and confirming the alternate hypothesis, H1. The results show that Neo4j is the fastest at executing queries, followed by MySQL and ArangoDB.

Similarly, Alyas et al. (2023) and Batra and Tyagi (2012) found that Neo4j performs better in processing connected data. The results also show that MySQL used more memory than Neo4j and ArangoDB. Finally, MySQL had higher energy consumption on average. However, contrary to Tsirogiannis et al. (2010) and Xu et al. (2012), it was insignificant. Future research could include more database technologies with simple to complex recursive queries and more datasets to make the results more generalizable. Other aspects that could be researched are different hardware configurations, storage, and how distributed database technologies perform since it is not sure that the current results are generalizable due to distributed computing environments.

# 7. References

Alyas, T., Alzahrani, A., Alsaawy, Y., Alissa, K., Abbas, Q., and Tabassum, N. (2023). Query Optimization Framework for Graph Database in Cloud Dew Environment. In *Computers, Materials & Continua*, 74(1), pp. 2317–2330. doi:10.32604/cmc.2023.032454.

ArangoDB Inc. (2023). *Roadmap*. Available at: https://www.arangodb.com/roadmap/ (Accessed: 10 March 2023).

Ayele, W. Y. and Juell-Skielse, G. (2017). Social media analytics and internet of things: survey. In Proceedings of the 1st International Conference on Internet of Things and Machine Learning, pp. 1-11. doi.org/10.1145/3109761.3158379.

Batra, S. and Tyagi, C. (2012). Comparative Analysis of Relational And Graph Databases. In *International Journal of Soft Computing and Engineering*, 2(2), pp. 509-512.

Bellini, P. and Nesi, P. (2018). Performance assessment of RDF graph databases for smart city services. In *Journal of Visual Languages & Computing, 45*, pp. 24-38. doi.org/10.1016/j.jvlc.2018.03.002.

BenchAnt. (2023). *This is benchANT*. Available at: https://benchant.com/company/about-us (Accessed: 14 February 2023).

Borg, E. and Westerlund, J. (2012). *Statistik för beteendevetare*. 3rd ed. Malmö (Sweden): Liber AB.

Cheng, Y., Ding, P., Wang, T., Lu, W., and Du, X. (2019). Which category is better: benchmarking relational and graph database management systems. In *Data Science and Engineering*, *4*, pp. 309-322.

Das, A., Mitra, A., Bhagat, S. N., and Paul, S. (2020). Issues and Concepts of Graph Database and a Comparative Analysis on list of Graph Database tools. In *2020 International Conference on Computer Communication and Informatics*. Coimbatore, India 22-24 January 2020, pp. 1–6. doi: 10.1109/ICCCI48352.2020.9104202.

Delacre, M., Leys, C., Mora, Y. L., and Lakens, D. (2019). Taking parametric assumptions seriously: Arguments for the use of Welch's F-test instead of the classical F-test in one-way ANOVA. In *International Review of Social Psychology*, *32*(1).

Denscombe, M. (2014). *The Good Research Guide For small-scale social research projects.* 5th ed. Berkshire (England): Open University Press.

Ellis, P. D. (2010). *The Essential Guide to Effect Sizes: An Introduction to Statistical Power, Meta-Analysis and the Interpretation of Research Results*. United Kingdom: Cambridge University Press.

Fernandes, D. and Bernardino, J. (2018). Graph Databases Comparison: AllegroGraph, ArangoDB, InfiniteGraph, Neo4J, and OrientDB. In *Proceedings of the 7th International Conference on Data Science, Technology and Applications (DATA 2018)*. Porto, Portugal 26 - 28 July, pp. 373-380. doi: 10.5220/0006910203730380.

George, D. and Mallery, M. (2010). *SPSS for Windows Step by Step: A Simple Guide and Reference, 17.0 update.* (10a ed.) Boston: Pearson.

Hair, J., Black, W. C., Babin, B. J. and Anderson, R. E. (2010). *Multivariate data analysis.* (7th ed.). Upper Saddle River, New Jersey: Pearson Educational International.

Hae-Young, K. (2014). Analysis of variance (ANOVA) comparing means of more than two groups, In *The Korean Academy of Conservative Dentistry, 39(1),* pp. 74-77. doi: 10.5395/rde.2014.39.1.74.

Hogan, A., Blomqvist, E., Cochez, M., d'Amato, C., Melo, G. D., Gutierrez, C., Kirrane, S., Gayo, J.E.L., Navigli, R., Neumaier, S., Ngomo, A.N., Polleres, A., Rashid, S.M., Rula, A., Schmelzeisen, L., Sequeda, J., Staab, S., and Zimmermann, A. (2021). Knowledge graphs. In *ACM Computing Surveys (CSUR), 54(4)*, pp. 1-37. doi: 10.1145/3447772.

Johannesson, P. and Perjons, E. (2014). *An Introduction to Design Science*. 1st ed. Cham (Switzerland): Springer International Publishing. doi: 10.1007/978-3-319-10632-8.

Jouili, S. and Vansteenberghe, V. (2013). An Empirical Comparison of Graph Databases. In *2013 International Conference on Social Computing.* Alexandria, VA, USA 8-14 September 2013, pp. 708-715. doi: 10.1109/SocialCom.2013.106.

Kaliyar, R. K. (2015). Graph databases: A survey. In *International Conference on Computing, Communication & Automation*. Greater Noida, India 15-16 May 2015, pp. 785-790. doi: 10.1109/CCAA.2015.7148480.

Li, Y. & Manoharan, S. (2013). A performance comparison of SQL and NoSQL databases. In *2013 IEEE Pacific Rim Conference on Communications, Computers and Signal Processing (PACRIM).* Victoria, BC, Canada, 27-29 August 2013, pp. 15-19. doi: 10.1109/PACRIM.2013.6625441.

McColl, R. C., Ediger, D., Poovey, J., Campbell, D., and Bader, D. A. (2014). A performance evaluation of open source graph databases. In *Proceedings of the first workshop on Parallel programming for analytics applications.* Orlando, Florida, USA 16 February 2014, pp. 11-18.

Mpinda, S. A. T., Ferreira, L. C., Ribeiro, M. X., and Santos, M. T. P. (2015). Evaluation of graph databases performance through indexing techniques. In *International Journal of Artificial Intelligence & Applications (IJAIA), 6(5)*, pp. 87-98. doi: 10.5121/ijaia.2015.6506.

Neo4j, (2023). *Neo4j Release Notes Archive: Database.* Available at: https://neo4j.com/release-notes/database/ (Accessed: 10 March 2023).

National Institute of Standards and Technology. (2012). Bartl*ett's Test*. Available at: https://www.itl.nist.gov/div898/handbook/eda/section3/eda357.htm (Accessed: 24 May 2023).

Novabench Inc. (2023). *Introduction Getting started with Novabench*. Available at: https://novabench.com/docs (Accessed: 7 February 2023).

Open Graph Benchmark. (2023). OGB Dataset Overview. Available at: https://ogb.stanford.edu/docs/dataset_overview/ (Accessed: 14 February 2023).

Oracle. (2023a). *MySQL 5.7 Release Notes.* Available at: https://dev.mysql.com/doc/relnotes/mysql/5.7/en/ (Accessed: 10 March 2023).

Oracle. (2023b). *MySQL 8.0 Release Notes.* Available at: https://dev.mysql.com/doc/relnotes/mysql/8.0/en/ (Accessed: 10 March 2023).

Ostertagova, E. and Ostertag, O. (2013). Methodology and Application of One-way ANOVA. In *American Journal of Mechanical Engineering, 1*, pp. 256-261.

Oxford Learner's Dictionary. (2023). *Benchmark*. Oxford University Press, Oxford. Available at: https://www.oxfordlearnersdictionaries.com/definition/english/benchmark_1?q=benchmark (Accessed: 7 February 2023).

Phaneendra, S. V., and Reddy, E. M. (2013). Big Data-solutions for RDBMS problems-A survey. In *12th IEEE/IFIP Network Operations & Management Symposium (NOMS 2010), 2(9),* pp. 3686-3691.

Piriform Software Ltd. (2023). Speccy. Available at: https://www.ccleaner.com/speccy (Accessed: 24 February 2023).

Pokorný, J. (2015). Graph Databases: Their Power and Limitations. In Saeed, K., Homenda, W. (eds.) Computer Information Systems and Industrial Management. Springer, pp 58 - 69. doi: 10.1007/978-3-319-24369-6_5.

Pokorný, J. (2016). Conceptual and database modelling of graph databases. In *Proceedings of the 20th international database engineering & applications symposium*, pp. 370-377. doi.org/10.1145/2938503.2938547.

Robinson, I., Webber, J., and Eifrem, E. (2015). *Graph Databases: New Opportunities for Connected Data*. 2nd ed. Sebastopol, CA: O'Reilly Media, Inc.

Ross, A., & Willson, V. L. (2017). One-way anova. In *Basic and advanced statistical tests* (pp. 21-24). Brill.

Sabolski, I., Leventić, H., and Galić, I. (2014). Performance evaluation of virtualization tools in multi-threaded applications. In *International journal of electrical and computer engineering systems, 5(2),* pp. 57-62.

SAS Whitepaper. (2012). Big Data Meets Big Data Analytics: Three Key Technologies for Extracting Real-Time Business Value from the Big Data That Threatens to Overwhelm Traditional Computing Architectures. SAS Insititue Inc., SAS Campus Drive Cary, NC.

Tsirogiannis, D., Harizopoulos, S., and Shah, M. A. (2010). Analyzing the energy efficiency of a database server. In *Proceedings of the 2010 ACM SIGMOD International Conference on Management of data*. Indianapolis, Indiana, USA, 6-11 June, pp. 231-242. doi.org/10.1145/1807167.1807194.

Ueta, K., Xue, X., Nakamoto, Y. and Murakami, S. (2016) A Distributed Graph Database for the Data Management

VanVoorhis, C. W., & Morgan, B. L. (2007). Understanding power and rules of thumb for determining sample sizes. *Tutorials in quantitative methods for psychology*, *3*(2), 43-50.

Xu, Z., Tu, Y. C., and Wang, X. (2012). PET: reducing database energy cost via query optimization. In *Proceedings of the VLDB Endowment, 5(12)*, pp. 1954-1957. doi.org/10.14778/2367502.2367546.

# Appendix A – database queries

**Query for Neo4j**

MATCH (n:SideEffect) ← [r:drug_sideeffect ] - (m:Drug) - [v:CURES] → (w:Disease) - [u:AFFECTS] → (p:Protein) - [:protein_catalysis] – (h:Protein) ← [:drug_protein] – (b:Drug) ← [:drug_sleep_disorder] - (m:Drug)
WHERE (m.drug_nr = 13)
RETURN DISTINCT n.name, m.name, w.name, p.protein_nr, b.drug_nr, h.protein_nr;

**Query for MySQL**

```
SELECT exjobb.drug.ent_name as "drugName", exjobb.disease.ent_name as "diseaseName",
exjobb.protein_new.ent_name as "proteinName", exjobb.drug_sleep_disorder_drug._to as
"sleepDisorderDrugID", exjobb.drug_to_protein._to as "sleepProteins",
exjobb.sideeffect.ent_name as "sideEffect"
FROM exjobb.protein_to_catalysis_protein, exjobb.drug_to_protein,
exjobb.drug_sleep_disorder_drug, exjobb.drug, exjobb.drug_to_sideffect, exjobb.sideeffect,
exjobb.drug_disease, exjobb.disease, exjobb.disease_protein, exjobb.protein_new
WHERE exjobb.drug.ent_idx = 13
AND exjobb.drug.ent_idx = exjobb.drug_to_sideffect._from
AND exjobb.sideeffect.ent_idx = exjobb.drug_to_sideffect._to
AND exjobb.drug.ent_idx = exjobb.drug_disease._from
AND exjobb.disease.ent_idx = exjobb.drug_disease._to
AND exjobb.disease.ent_idx = exjobb.disease_protein._from
AND exjobb.protein_new.ent_idx = exjobb.disease_protein._to
AND exjobb.protein_new.ent_idx = exjobb.protein_to_catalysis_protein._to
AND exjobb.drug_to_protein._to = exjobb.protein_to_catalysis_protein._from
AND exjobb.drug_to_protein._from = exjobb.drug_sleep_disorder_drug._to
AND exjobb.drug_sleep_disorder_drug._from = exjobb.drug.ent_idx
```

**Query for ArangoDB used in Arango Shell**

db._query('LET side = (for v,e,p IN 1..2 Outbound "Drugs/15355" Relation_Drug_To_SideEffect RETURN distinct {SideEffect: v}) LET disease = (for traversal in side for v1,e1,p1 IN 1..1 Outbound "Drugs/15355" Relation_Drug_To_Disease RETURN distinct {Disease: v1}) LET protein = (for traversal in disease for v2,e2,p2 IN 1..1 Outbound traversal.Disease Relation_Disease_to_Protein RETURN distinct { Protein: v2}) LET catalysis = (for traversal in protein for v3,e3,p3 IN 1..1 Outbound traversal.Protein Relation_Protein_To_Catalysis_Protein RETURN distinct {Catalysis : v3}) LET sleepDrugsSingleLinked= (for traversal in catalysis for v4,e4,p4 IN 1..1 Inbound traversal.Catalysis Relation_Drug_To_Protein RETURN distinct {SleepDisorderDrugSingleLinked: v4}) LET SleepDisorderDrug= (for traversal in sleepDrugsSingleLinked for v5,e5,p5 IN 1..1 Inbound traversal.SleepDisorderDrugSingleLinked Relation_Drug_Sleep_Disorder prune v5._id == "Drugs/15355" RETURN distinct {SleepDisorderDrug: v5}) RETURN {Drug: "Drugs/15355", SideEffects: side[*], Disease: disease[*], Protein: protein[*], Catalysis_Protein: catalysis[*], Sleep_Disorder_Drug: SleepDisorderDrug[*]}').getExtra();

**Readable format of the query for ArangoDB**

```
LET side = (
        for v,e,p IN 1..2 Outbound "Drugs/15355"
        Relation_Drug_To_SideEffect
```

```
                    RETURN distinct {SideEffect: v}
                )
LET disease = (
                for traversal in side
                for v1,e1,p1 IN 1..1 Outbound "Drugs/15355"
                Relation_Drug_To_Disease
                RETURN distinct {Disease: v1}
                )
LET protein = (
                for traversal in disease
                for v2,e2,p2 IN 1..1 Outbound traversal.Disease
        Relation_Disease_to_Protein
                RETURN distinct { Protein: v2}
                )
LET catalysis = (
                for traversal in protein
                for v3,e3,p3 IN 1..1 Outbound traversal.Protein
                Relation_Protein_To_Catalysis_Protein
                RETURN distinct {
                Catalysis : v3}
                )
LET sleepDrugsSingleLinked= (
                for traversal in catalysis
                for v4,e4,p4 IN 1..1 Inbound traversal.Catalysis
                Relation_Drug_To_Protein
                RETURN distinct {
                SleepDisorderDrugSingleLinked: v4}
                )
LET SleepDisorderDrug= (
                for traversal in sleepDrugsSingleLinked
                for v5,e5,p5 IN 1..1 Inbound traversal.SleepDisorderDrugSingleLinked
                Relation_Drug_Sleep_Disorder
                prune v5._id == "Drugs/15355"
                RETURN distinct {
                SleepDisorderDrug: v5}
                )
RETURN {Drug: "Drugs/15355", SideEffects: side[*], Disease: disease[*], Protein: protein[*],
Catalysis_Protein: catalysis[*], Sleep_Disorder_Drug: SleepDisorderDrug[*]}
```

## Appendix B – outliers

### ArangoDB

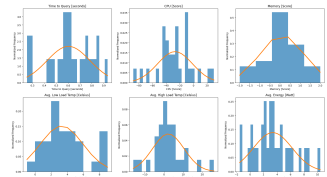

### MySQL

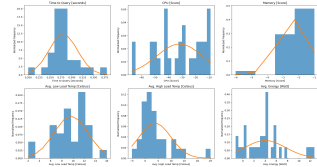

### Neo4j

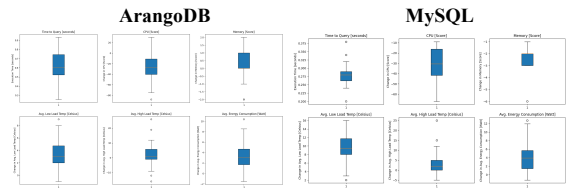

## Appendix C – symmetry, tail-heaviness, and assumption of Normality

### ArangoDB

| Aspect | Skewness | Kurtosis |
|---|---|---|
| Time to Query | -0.392241 | -0.293994 |
| CPU | -0.082038 | 0.658869 |
| Memory | -0.21001 | -0.140723 |
| Avg. Low Load Temperature | 0.649941 | -0.256214 |
| Avg. High Load Temperature | 0.60294 | 2.034986 |
| Avg. Energy Usage | 0.548555 | 0.199047 |

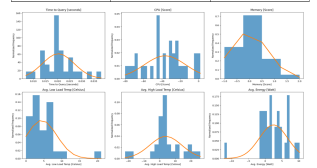

### MySQL

| Aspect | Skewness | Kurtosis |
|---|---|---|
| Time to Query | 0.682635 | 2.838409 |
| CPU | -0.39406 | -0.726857 |
| Memory | -1.512037 | 5.296763 |
| Avg. Low Load Temperature | -0.54087 | 0.886703 |
| Avg. High Load Temperature | 1.77043 | 4.214859 |
| Avg. Energy Usage | 0.850486 | 0.727137 |

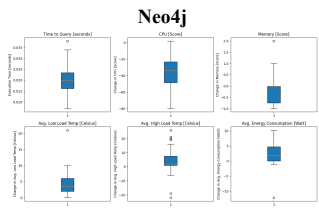

### Neo4j

| Aspect | Skewness | Kurtosis |
|---|---|---|
| Time to Query | 0.57783 | 1.226893 |
| CPU | -0.179711 | -0.637931 |
| Memory | 0.552966 | 0.353232 |
| Avg. Low Load Temperature | 2.160023 | 7.386958 |
| Avg. High Load Temperature | -0.199822 | 1.476351 |
| Avg. Energy Usage | -1.068917 | 3.936231 |